\documentclass[aps,prl,twocolumn,amsmath,amssymb,superscriptaddress]{revtex4-2}

\usepackage{graphicx} 

\newcommand{\LL}[1]{\textcolor{black}{#1}}

\usepackage[utf8]{inputenc}

\usepackage{lipsum}
\usepackage{svg}
\usepackage{graphicx}
\usepackage{braket}
\usepackage{tabularx}
\usepackage{gensymb}
\usepackage{color}

\usepackage{makecell}

\usepackage{comment}
\usepackage{lineno}

\begin{document}
	
	\title{\textbf{\LL{Measurement-induced back-action in a QD-based coherent spin-photon interface}}}
	
	
	
	
	
	
	
	
	\author{Adrià Medeiros}
	\affiliation{Université Paris-Saclay, CNRS, Centre de Nanosciences et de Nanotechnologies, 91120, Palaiseau, France
	}
	\author{M. Gund{\'i}n}
	\affiliation{Université Paris-Saclay, CNRS, Centre de Nanosciences et de Nanotechnologies, 91120, Palaiseau, France
	}
	\author{D. A. Fioretto}
	\affiliation{Université Paris-Saclay, CNRS, Centre de Nanosciences et de Nanotechnologies, 91120, Palaiseau, France
	}
	\author{V. Vinel}
	\affiliation{Université Paris-Saclay, CNRS, Centre de Nanosciences et de Nanotechnologies, 91120, Palaiseau, France
	}
	\author{E. Rambeau}
	\affiliation{Université Paris-Saclay, CNRS, Centre de Nanosciences et de Nanotechnologies, 91120, Palaiseau, France
	}
	\affiliation{Université Paris Cité, Centre de Nanosciences et de Nanotechnologies, 91120 Palaiseau, France} 
	\author{E. Mehdi}
	\affiliation{Université Paris-Saclay, CNRS, Centre de Nanosciences et de Nanotechnologies, 91120, Palaiseau, France
	}
	\author{N. Somaschi}
	\affiliation{Quandela, 7 rue Leonard de Vinci, 91300 Massy, France}
	\author{A. Lemaître}
	\affiliation{Université Paris-Saclay, CNRS, Centre de Nanosciences et de Nanotechnologies, 91120, Palaiseau, France
	}
	\author{I. Sagnes}
	\affiliation{Université Paris-Saclay, CNRS, Centre de Nanosciences et de Nanotechnologies, 91120, Palaiseau, France
	}
	\author{N. Belabas}
	\affiliation{Université Paris-Saclay, CNRS, Centre de Nanosciences et de Nanotechnologies, 91120, Palaiseau, France
	}
	\author{O. Krebs}
	\affiliation{Université Paris-Saclay, CNRS, Centre de Nanosciences et de Nanotechnologies, 91120, Palaiseau, France
	}
	\author{P. Senellart}
	\affiliation{Université Paris-Saclay, CNRS, Centre de Nanosciences et de Nanotechnologies, 91120, Palaiseau, France
	}
	\author{L. Lanco}
	\affiliation{Université Paris-Saclay, CNRS, Centre de Nanosciences et de Nanotechnologies, 91120, Palaiseau, France
	}
	\affiliation{Université Paris Cité, Centre de Nanosciences et de Nanotechnologies, 91120 Palaiseau, France} 
	
	\begin{abstract}
		Spin-photon interfaces constitute promising candidates for the development of stationary nodes used as photon receivers, for quantum communication and distributed quantum computing. Here we introduce a time-resolved tomography approach which allows observing the dynamics of an electron spin, in a semiconductor quantum dot, mapped onto the dynamics of the polarization state of reflected photons.
		Through a single tomography experiment, we infer all the relevant spin dynamics timescales, including  precession, decoherence and relaxation times. 
		We also demonstrate and quantify the measurement back-action induced, on the embedded spin qubit, by the detection of a single reflected photon. We show that the induced population and coherence of the spin state can be tuned by the chosen polarization basis of the measurement. 
		The control of the photon-induced back-action on the embedded spin qubit constitutes a crucial requirement for the use of spin-photon interfaces as quantum receivers.

	\end{abstract}
	
	\maketitle
	
	\section{INTRODUCTION}
	
	
	
	Efficient light-matter interfaces, allowing for the coherent transfer of quantum information between a single photon and a single atom \cite{Cirac1997}, hold great promise \LL{for} the development of photonic quantum information processing architectures \cite{Kimble2008, Reiserer2022}. Such interfaces could allow for potentially deterministic photon-photon gates \cite{Buterakos2017}, a key step towards the generation of highly-entangled photonic graph states for scalable measurement-based quantum computing  \cite{Briegel2009, Raussendorf2007}.
	
	
	Complementarily to the development of efficient quantum emitters \cite{Couteau2023, Thomas2022, Coste2023_Cluster}
	, a number of systems have been explored as coherent receivers, \LL{controllably modifying the state of incoming photons following their interaction} with atoms \cite{Tiecke2014,Volz2014,Bechler2018,Daiss2021,Stolz2022}, superconducting quantum circuits \cite{Ferreira2024, Wang2022,Reuer2022}, semiconductor quantum dots \cite{DeSantis2017,LeJeannic2022,Hansen2024,Mehdi2024}, vacancy centers in diamond \cite{Bhaskar2020,Pasini2024,Herrmann2024} \LL{or} rare-earth ions \cite{Zhong2015}. In this context, solid-state spins \cite{Atature2018, Awschalom2018} have emerged as promising candidates for the development of scalable devices which can be operated at high rates, if efficiently coupled to optical microcavities \cite{Coste2023_Cluster} or nanophotonic waveguides \cite{Bhaskar2020}. In the optical domain, spin-photon interfaces \cite{Borregaard2019} were used as coherent receivers mainly with color centers in diamond and semiconductor quantum dots, for quantum memory applications \cite{Nguyen2019_Nodes, Stas2022}, remote spin-photon entanglement \cite{Bhaskar2020, Knaut2024, Chan2023} and spin-photon gates \cite{Parker2024, Sun2016, Sun2018}.

	In particular, semiconductor quantum dots (QDs) embedded in  micropillar cavities \cite{Arnold2015,Wells2019,Androvitsaneas2019,Mehdi2024,Androvitsaneas2024,Gundin2025} hold great promise due to the presence of giant spin-dependent Kerr rotations \cite{Arnold2015, Mehdi2024}, where the polarization of a reflected photon depends on the resident spin state. In the last decade, a number of theoretical proposals exploiting this effect have emerged for fault-tolerant quantum computing \cite{Leuenberger2006}, deterministic logic gates \cite{Koshino2010} or spin-mediated photon-photon entanglement \cite{Hu2008_Entangler, Hu2008}.
	
	From a fundamental perspective, spin-dependent \LL{Kerr} rotations can be used to probe the QD-cavity system through measurement-induced back-action, whereby the spin quantum state is modified after the detection of a single reflected photon \cite{Maffei2023, Smirnov2017}. It was theoretically shown \LL{that}, using a spin-photon interface in the strong-coupling regime under a transverse magnetic field, the spin can be \LL{projected out of equilibrium }after an initial detection event, and \LL{its} subsequent dynamics transferred onto the transmitted light intensity 
	\cite{Smirnov2017}. The photon-induced back-action can be used to produce coherent superpositions of spin eigenstates, a crucial resource to \LL{subsequently implement} entangling gates with incoming photons \cite{Hu2008,Hu2008_Entangler,Lindner2009}.
	
	In a previous experiment, the measurement back-action was used to probe the relaxation of a single hole spin\LL{, through giant Kerr rotations}, under a longitudinal magnetic field \cite{Gundin2025}. In such a \LL{magnetic field configuration}, detecting a single \LL{reflected} photon is equivalent to a classical measurement: it only heralds a spin population imbalance, in the basis of the spin eigenstates, with no quantum coherence involved. Up to now, there have been been no theoretical nor experimental studies  \LL{demonstrating the possibility of inducing a truly quantum back-action, producing coherent spin superpositions, using realistic interfaces}. The possibility of mapping a spin-qubit precession to a polarization-qubit precession\LL{, within the limits permitted by the no-cloning theorem \cite{Wootters1982},} has also been left unexplored.
	
	\begin{figure*}[t!]
		\includegraphics[scale=1]{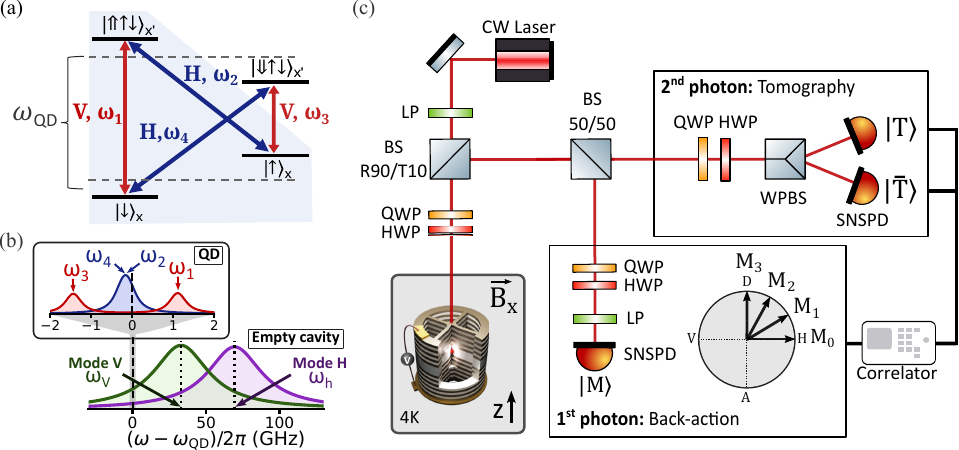}
		\caption{(a) Energy levels of a negatively charged QD under a transverse external magnetic field. We excite the $\Lambda$-system shown in the shaded area by driving transition $(V,\ \omega_1)$ with a resonant CW laser.
			(b) Numerically simulated empty cavity spectrum (bottom) and QD transition energies (top) as a function of the detuning with the QD transition frequency at 0T (dashed vertical line).
			(c) Experimental setup for studying the dynamics of the QD spin. A first reflected photon is measured in a given linear polarization $M$ on a calibrated polarimeter (bottom right box, the inset shows the four experimentally-used polarization states) to induce a back-action on the spin. A second polarimeter (top right) performs the tomography of a second reflected photon state, conditioned to the first photon detection. BS: Beamsplitter, QWP: Quarter Wave Plate, HWP: Half Wave Plate, LP: Linear Polarizer, WPBS: Wollaston Polarizing Beamsplitter, SNSPD: Super-conducting Nanowire Single Photon Detectors.
		}
		\centering
	\end{figure*}
	
	In the present work, we exploit cavity-enhanced Kerr rotations to demonstrate the quantum back-action induced by the detection of a single reflected photon, and the mapping between the spin state dynamics and the polarization state dynamics of reflected photons. \LL{To do so we introduce a new experimental approach based on time-resolved polarization tomography measurements, exploiting the cavity-enhanced Kerr rotation in a pillar-based spin-photon interface}. We use the continous-wave (CW) excitation of a charged QD-cavity device, in a weak transverse magnetic field, to project an electron spin state through the detection of a single reflected photon. We find that the modified spin state, after back-action, acquires some coherence in the basis of the energy eigenstates, and can thus precess around the applied magnetic field. \LL{After the initial back-action, t}he spin evolution towards equilibrium is tracked by performing the polarization tomography of \LL{a subsequent reflected photon, through 2-photon correlation measurements performed in} adequately chosen polarization bases. We introduce an analytical model describing the measurement back-action as a consequence of the interference between the QD resonant fluorescence and the light directly reflected by the empty cavity. This allows explaining how the three Bloch components of the spin state can be partially mapped to the three Stokes components of the photon polarisation state. 
	The experimental data are complemented with a full numerical description of the system, which confirms that one can infer the complete spin dynamics, including the Larmor frequency $\omega_\mathrm{L}$, the spin coherence time $T_\mathrm{2,spin}^*$ and the spin relaxation time $T_\mathrm{1,spin}$, from a single \LL{time-resolved, conditional}
	tomography measurement. Finally, we show how the photon-induced back-action can be maximized through the control of the measured polarization axis. This new experimental approach provides \LL{useful insights and preliminary steps towards}  the practical implementation of  spin-photon entanglement and logic gates, using single photons incoming to and reflected from pillar-based spin-photon interfaces. \\

	\section{RESULTS}
	\subsection{QD-micropillar cavity device}
	The device under study is a negatively-charged InGaAs QD emitting at 925 nm, embedded in an AlGaAs micropillar cavity \cite{Somaschi2016} (see \LL{Appendix}). A transverse magnetic field (Voigt configuration), at $B_x=200\rm~mT$, is applied perpendicularly to the pillar symmetry axis representing the $z$ direction. In Fig.~1a we display the QD energy levels, shifted by the Zeeman effect. The two electron spin states $\ket{\uparrow}_x$ and $\ket{\downarrow}_x$ form the ground states. The excited states are the two trion states $\ket{\Uparrow\uparrow\downarrow}_{x'}$ and $\ket{\Downarrow\downarrow\uparrow}_{x'}$, consisting of a pair of electrons in a singlet configuration and a single hole, whose  spin eigenstates are generally not oriented along the magnetic field axis $x$ \cite{Ramesh2025}.
	The states are coupled to light with two orthogonal linear polarizations ($V$ and $H$ respectively, with subscript $i=1,2,3,4$ used to label each transition energy $\omega_i$) \cite{Warburton2013}. In our experiment, we resonantly excite transition ($V, \omega_1$), operating in a lambda configuration coupling the two ground states $\ket{\uparrow}_x$ and $\ket{\downarrow}_x$ to the excited state $\ket{\Uparrow\downarrow\uparrow}_{x'}$. Using a minimally invasive CW laser in the low-power regime ($\mathrm{P}\sim 6\ \mathrm{pW}$), to avoid optical spin pumping \cite{Xu2007}, allows maintaining a steady state close to the ground-state spin thermal equilibrium. 
	
	The optical properties of the micropillar cavity, extracted from polarization-dependent reflectivity measurements \cite{Anton2017,Hilaire2018} are illustrated in Fig. 1b. The fundamental cavity mode is split into two orthogonal linear polarization modes of resonance energies $\omega_{\rm H}$ and $\omega_{\rm V}$ split by $\Delta_{\rm c}=\omega_{\rm H}-\omega_{\rm V}=2\pi\times(36.4\pm0.7)\rm~GHz$. The cavity polarization eigenaxes \LL{are determined by the anisotropy of the internal strain field in the micropillar device. Such eigenaxes actually match the optical transitions' polarizations $H$ and $V$ in Fig~1a, which are governed by the anisotropy of the transverse hole g-tensor, the latter being mostly determined by the internal strain field anisotropy \cite{Ramesh2025}}. 
	The cavity damping rates are $\kappa_H/2\pi = 44.5 \pm  0.4~\rm GHz$ and $\kappa_V/2\pi = 45.0 \pm 0.4~\rm GHz$. The effective top mirror output coupling for the $V$($H$)-polarized mode is $\eta_{\mathrm{top}, V} = 0.63 \pm 0.01$ ($\eta_{\mathrm{top}, H} = 0.65 \pm 0.01$) \cite{Mehdi2024}. The QD resonance frequency at 0T, $\omega_{\rm QD}$, is red-detuned from the center cavity frequency $\omega_c=\left(\omega_H + \omega_V\right)/2$ by \mbox{$\Delta_{\rm QD}=\omega_{\rm QD} - \omega_c = -2\pi\times(51.2\pm 0.7)\ \mathrm{GHz}$}. This undesired detuning diminishes the device performance but remains small enough, compared to the cavity damping rates, to allow a significant coupling of the optical transitions to the cavity modes.\\
	
	The optical setup for this experiment is schematized in Fig.~1c. The polarization of the input light, at frequency $\omega_1$, is fixed with a linear polarizer (LP) and modified with a set of half- and quarter- waveplates (HWP/QWP) to excite only the cavity mode $V$ and, thus, the transition $(V,\omega_1)$. The laser is focused into the micropillar cavity within a cryostat operating at 4K \cite{Mehdi2024}. The reflected light is separated by a R90/T10 non-polarizing beam splitter (BS) and spatially filtered in a single mode fiber (not shown). A 50/50 free space BS divides the signal into two polarization analyzers.
	The first polarization analyzer (bottom right) measures a first reflected photon in a linear polarization state parametrized by the angle $\phi$:
	\begin{equation}
		\ket M=\cos(\phi/2)\ket H_{\rm refl}+\sin(\phi/2)\ket V_{\rm refl},
		\label{eq:Polarization_parametrization}
	\end{equation}leading to a back-action on the spin. For this experiment, we measure four different polarization states labeled as $M_0$, $M_1$, $M_2$ and\ $M_3$ described by the angle $\phi = \{0, \frac{\pi}{6}, \frac{\pi}{3} \mathrm{\ and\ } \frac{\pi}{2}\}$, where $\phi=0$ corresponds to polarization $H$ and $\phi = \pi/2$ to the diagonal polarization $D$. The second analyzer (top right) performs the polarization tomography of a second reflected photon by measuring in three orthogonal bases $T/\bar T=H/V,~D/A,~R/L$.
	
	We note that such a technique extends the approach experimentally pioneered in Ref. \cite{Coste2023,Serov2024}, where pillar-based interfaces were used only as photon emitters and were studied only through cross-correlations, in orthogonal measurements bases, e.g. $H/V$ or $R/L$. A key ingredient of our approach lies in the possibility to independently choose the polarization axis for the first detected photon, to control the measurement-induced back-action on the spin state, and for the subsequent reflected photon after a delay $\tau$, to probe the effect of this back-action through polarization tomography.\\
	
	\subsection{Analytical model of the back-action}
	We begin by introducing an analytical model describing the physical mechanism behind the measurement back-action explored in this work. Initially, i.e. before any photon has been detected,  the spin is in a statistical mixture close to thermal equilibrium, defined by the state:
	\begin{equation}
		\rho^{\mathrm{(s)}}_0 = P_\uparrow\ket{\uparrow}\bra{\uparrow} + P_\downarrow\ket{\downarrow}\bra{\downarrow},
		\label{eq:original_mix_state}
	\end{equation}
	where $\rho_0^{\mathrm{(s)}}$ refers to the spin density matrix and $P_\uparrow\ (P_\downarrow)$ is the probability of finding the spin in $\ket{\uparrow(\downarrow)}_x$. From now on, we will ignore the $x$ label for the spin states. When an incoming photon, resonant and co-polarized with transition $(V, \omega_1)$,  interacts with the QD-cavity device, the resulting spin-scattered photon state can be written as:
	\begin{equation}
		\rho^{\rm (s,p)} = \sum_{i=\uparrow, \downarrow} P_i\ket{\Psi^{\mathrm{(s,p)}}_i}\bra{\Psi^{\mathrm{(s,p)}}_i},
		\label{interaction_description_mixed_state}
	\end{equation}
	where $\ket{\Psi^{\mathrm{(s,p)}}_\uparrow}$ (resp. $\ket{\Psi^{\mathrm{(s,p)}}_\downarrow)}$) is the spin-scattered photon state obtained for a spin initially in state $\ket{\uparrow}$ (resp. $\ket{\downarrow}$):
	\begin{equation}
		\begin{split}
			\ket{\Psi^{\mathrm{(s,p)}}_{\uparrow}} & =  r_{\uparrow\uparrow}\ket{\uparrow} \ket{V,\omega_1}_{\mathrm{refl}} + ...\\
			\ket{\Psi^{\mathrm{(s,p)}}_{\downarrow}} & =  r_{\downarrow\downarrow}\ket{\downarrow} \ket{V,\omega_1}_{\mathrm{refl}} + r_{\downarrow\uparrow}\ket{\uparrow} \ket{H,\omega_2}_{\mathrm{refl}} + ...
		\end{split}
		\label{interaction_description}
	\end{equation}
	The coefficients $r_{ij}$ (with $i,\ j$ either $\uparrow$ or $\downarrow$) are reflection amplitudes associated to spin-preserving and frequency preserving transitions ($r_{\uparrow\uparrow},\ r_{\downarrow\downarrow}$) or Raman spin-flip transitions ($r_{\downarrow\uparrow}$). The term $\ket{\Psi^{\mathrm{(s,p)}}_\uparrow}$ describes the spin-scattered photon state obtained when the spin is initially in $\ket{\uparrow}$. In such a case, the QD remains unexcited by the CW laser which only allows exciting the $(V,\omega_1)$ transition. The reflection amplitude $r_{\uparrow\uparrow}$ then corresponds to the empty cavity response, and we take $r_{\uparrow\downarrow}=0$ (\LL{neglecting the probability of a} Raman-spin flip transition from $\ket{\uparrow}$ to $\ket{\downarrow}$\LL{, which would require exciting at a  frequency closer to $\omega_3$)}. 
	Conversely, $\ket{\Psi^{\mathrm{(s,p)}}_\downarrow}$ describes the spin-scattered photon state when the spin is initially in state $\ket{\downarrow}$. 
	In this case, $r_{\downarrow\downarrow}$ is governed by the interference between the directly-reflected light and the co-polarized Rayleigh scattered emission from the QD, both at frequency $\omega_1$. The reflection amplitude $r_{\downarrow\uparrow}$ results from the cross-polarized Raman emission at frequency $\omega_2$ leading to a spin-flip from $\ket{\downarrow}$ to $\ket{\uparrow}$.
	The remaining terms relate to other scattering channels (such as transmitted, diffracted or lost light, or light spontaneously emitted outside the cavity modes) which we do not require to take into account in our experiment, as we post-select on the detection of a reflected photon.
	
	We now consider the case where a scattered photon has been reflected and detected in the linear polarization state $\ket M$ (see Eq.~\ref{eq:Polarization_parametrization}). The detection of a reflected photon in polarization $\ket M$ leads to a back-action on the spin system, whose modified density matrix is described by \mbox{$\rho^{\mathrm{(s)}}_{|M}=\braket{M|\rho^{\rm (s,p)}|M}/\mathrm{Tr}\left(\braket{M|\rho^{\rm (s,p)}|M}\right)$}, with components:
	\begin{equation}
		\begin{split}
			\rho^{\mathrm{(s)}}_{|M, \downarrow\downarrow} &=\frac{P_\downarrow}{P_M(\phi)}|r_{\downarrow\downarrow}|^2\sin^2{(\phi/2)}\\
			\rho^{\mathrm{(s)}}_{|M, \downarrow\uparrow} &=\frac{P_\downarrow}{2P_M(\phi)}r_{\downarrow\downarrow}r^*_{\downarrow\uparrow}\sin{\left(\phi\right)}, 
		\end{split}
		\label{eq:state_after_measurement}
	\end{equation}
	as well as $\rho^{\mathrm{(s)}}_{|M,\uparrow\uparrow}=1-\rho^{\mathrm{(s)}}_{|M,\downarrow\downarrow}$ and $\rho^{\mathrm{(s)}}_{|M, \downarrow\uparrow}~=~\left(\rho^{\mathrm{(s)}}_{|M, \downarrow\uparrow}\right)^*$. In these expressions, $P_M(\phi)$ represents the probability of actually detecting an $M$-polarized reflected photon:
	\begin{equation}
		\begin{split}
			P_M(\phi) = &P_\uparrow|r_{\uparrow\uparrow}|^2\sin^2(\phi/2)+\\
			&P_\downarrow\left(|r_{\downarrow\uparrow}|^2\cos^2(\phi/2)+|r_{\downarrow\downarrow}|^2\sin^2(\phi/2)\right)
		\end{split}
		\label{eq:P_M-probability}
	\end{equation}
	\begin{figure*}[t!]
		\includegraphics[scale=1]{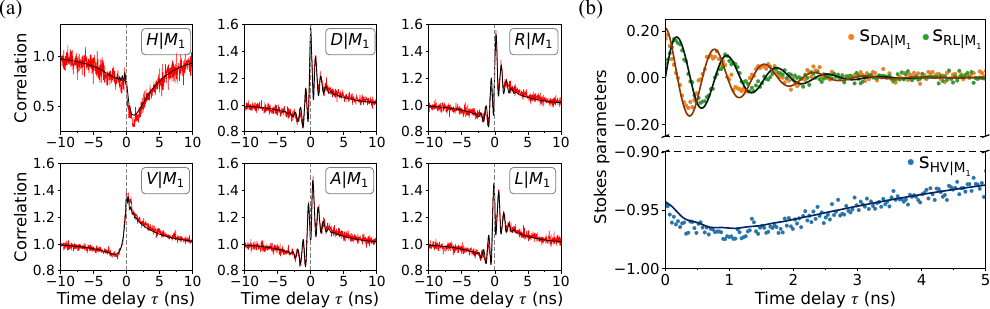}
		\caption{(a) Two-photon time-correlation measurements between a photon in the projection state $M_1$ and another one in each of the six states defining a set of orthogonal bases of the Poincaré sphere.  Correlations \LL{$H|M_1$ and $V|M_1$} capture the change in the populations of the projected spin state, while the remaining four also display signatures of measurement-induced coherence. 
			(b) Time-evolution of the conditional Stokes parameters \LL{as a function of the delay $\tau$ after the detection of a photon with polarization $M_1$}. \LL{In all panels }the scatter plot display the experimental data and the black solid lines correspond to numerical simulations.
		}
		\centering
	\end{figure*}
	The case $\phi=0$ corresponds to the measurement of an $H$-polarized photon, coming from a Raman spin-flip transition, with probability $P_H=P_M(0)=P_\downarrow|r_{\downarrow\uparrow}|^2$,  projecting the spin in the $\ket{\uparrow}$ state. This \LL{leads to conditional spin state} populations $\rho_{|H,\uparrow\uparrow}^{\mathrm{(s)}}=1$ and $\rho_{|H,\downarrow\downarrow}^{\mathrm{(s)}}=0$, with no coherence i.e. $\rho_{|H,\downarrow\uparrow}^{\mathrm{(s)}}=0$. The case $\phi=\pi$ corresponds to the measurement of a $V$-polarized photon, with probability $P_V=P_M(\pi)=P_\uparrow|r_{\uparrow\uparrow}|^2+P_\downarrow|r_{\downarrow\downarrow}|^2$, which partially projects the spin such that $\rho_{|V,\uparrow\uparrow}^{\mathrm{(s)}}=P_\downarrow|r_{\downarrow\downarrow}|^2/(P_\uparrow|r_{\uparrow\uparrow}|^2+P_\downarrow|r_{\downarrow\downarrow}|^2)$  with no coherence ($\rho_{|V,\downarrow\uparrow}^{\mathrm{(s)}}~=~0$). 
	
	In both these situations, the back-action onto the spin state through the detection of a reflected photon acts only on the populations. To induce some coherence one needs to exploit the entanglement present in the $\ket{\Psi^{\mathrm{(s,p)}}_{\downarrow}}$ state, through the measurement of a polarization state being in a superposition of $\ket{H}$ and $\ket{V}$, i.e. $\phi\neq\{0, \pi\}$. This is seen in Eq. (\ref{eq:state_after_measurement}), where the coherence term is both proportional to $\sin(\phi)$ and to the product $r_{\downarrow\downarrow}r_{\downarrow\uparrow}^*$.
	
	After the projective measurement, the conditional density matrix $\rho_{|M}^{\mathrm{(s)}}$ evolves towards the steady state (Eq.~(\ref{eq:original_mix_state})) due to spin relaxation (affecting the populations $\rho_{|M,\downarrow\downarrow}^{\mathrm{(s)}}$ and $\rho_{|M,\uparrow\uparrow}^{\mathrm{(s)}}$) and decoherence (affecting the coherence $\rho_{|M,\downarrow\uparrow}^{\mathrm{(s)}}$). One can track this dynamics by performing the polarization tomography of a second reflected photon, using measurements in three orthogonal bases ($T\bar{T}=HV, DA, RL$), at a time delay $\tau$ after the first photon is detected in polarization $M$. This allows measuring the conditional Stokes parameters:
	\begin{equation}
		s_{T\bar{T}|M}=\frac{P_{T|M}(\tau) - P_{\bar{T}|M}(\tau)}{P_{T|M}(\tau) + P_{\bar{T}|M}(\tau)},
		\label{eq:definition_conditional_stokes_parameters}
	\end{equation}
	where $P_{T(\bar{T})|M}(\tau)$ is the probability of detecting a photon in the polarization state $T(\bar{T})$ at a time delay $\tau$ conditioned to the detection of a first photon in polarization $M$ at $\tau=0$. In particular, the conditional Stokes parameters $s_{HV|M}$, $s_{DA|M}$ and $s_{RL|M}$ are related to either the populations or the coherence of the spin state $\rho^{\mathrm{(s)}}_{|M}(\tau)$ through:
	\fontsize{9pt}{12pt}\selectfont
	\begin{equation}
		\begin{split}
			s_{HV|M} &= \frac{(|r_{\downarrow\uparrow}|^2-|r_{\downarrow\downarrow}|^2)\rho^{\mathrm{(s)}}_{|M,\downarrow\downarrow}(\tau)-|r_{\uparrow\uparrow}|^2\rho^{\mathrm{(s)}}_{|M,\uparrow\uparrow}(\tau)}{P_{\mathrm{refl}|M}(\tau)}\\
			s_{DA|M} &= \frac{2\times\mathrm{Re}\left(r_{\uparrow\uparrow}r^*_{\downarrow\uparrow}\rho^{\mathrm{(s)}}_{|M,\uparrow\downarrow}(\tau)\right)}{P_{\mathrm{refl}|M}(\tau)}\\
			s_{RL|M} &= \frac{2\times\mathrm{Im}\left(r_{\uparrow\uparrow}r^*_{\downarrow\uparrow}\rho^{\mathrm{(s)}}_{|M,\uparrow\downarrow}(\tau)\right)}{P_{\mathrm{refl}|M}(\tau)},
			\label{eq:conditional_stokes_parameters}
		\end{split}
	\end{equation}
	
	\normalsize
	with:
	\fontsize{9pt}{12pt}\selectfont
	\begin{equation}
		P_{\mathrm{refl}|M}(\tau)=(|r_{\downarrow\uparrow}|^2+|r_{\downarrow\downarrow}|^2)\rho^{\mathrm{(s)}}_{|M,\downarrow\downarrow}(\tau)+
		|r_{\uparrow\uparrow}|^2\rho^{\mathrm{(s)}}_{|M,\uparrow\uparrow}(\tau)
		\label{eq:P_refl_M}
	\end{equation}
	\normalsize
	the total reflection probability for a second photon, conditioned to the detection of a first photon in polarisation $M$. Such formulae show that the modified spin state is mapped to the polarization state of the next reflected photon. This mapping does not constitute a bijection, which would break the no-cloning theorem \cite{Wootters1982}. Still, it leads to a unique and direct correspondance between the spin density matrix and the reflected photon's density matrix, governed by the three Stokes parameters in Eq.~(\ref{eq:conditional_stokes_parameters}).
	
	Through this model, one can qualitatively predict the expected evolution of the conditional Stokes parameters after the initial measurement back-action induced at $\tau=0$ by the detection of an $M$-polarized photon. The parameters $s_{DA|M}(\tau)$ and $s_{RL|M}(\tau)$ should display damped oscillations in quadrature, as the non-diagonal coherence term $\rho_{|M,\uparrow\downarrow}$ evolves in phase at the Larmor frequency $\omega_L$, and eventually decoheres with an effective rate $T_{2,\rm spin}^*$. Conversely, $s_{HV|M}(\tau)$ should decay towards its stationary value at a rate $T_{1,\rm spin}$, following the relaxation of the conditional populations towards $P_\uparrow$ and $P_\downarrow$.\\
	
	\subsection{Polarization dynamics after spin projection}
	The conditional probabilities $P_{T|M}(\tau)$ defined in Eq.~\ref{eq:definition_conditional_stokes_parameters} are accessed through time-resolved 2-photon correlations:
	\begin{equation}
		g^{(2)}_{T|M}(\tau)=\frac{P_{T|M}(\tau)}{P_T},
	\end{equation}
	where a first photon is experimentally measured in polarization state $\ket M$ and the second in polarization state $\ket T$ (see Fig.~1c), with $P_T$ the uncorrelated probability of detecting a $T$-polarized photon. \LL{To discuss the quantum back-action phenomenon}, we first display in Fig.~2a the measured \LL{correlations} with the first photon measured in polarization $M_1~(\phi=30\degree)$. In this figure, both for positive and negative delays, the experimental data are found to be in excellent agreement with numerical simulations discussed later on (see Supplemental Materials, hereafter denoted as Ref. \cite{SM}, for complete model discussion). In the following, we focus on the correlations at positive delays which correspond to the previously-discussed situation, where the $M_1$-polarized photon is detected before the $T$-polarized one. Negative delays correspond to the complementary situation, in which the measurement of a $T$-polarized photon is detected before the $M_1$-polarized one.
	
	When $T=H/V$, no coherent oscillations are observed, as the detection of an $H$ or $V$ polarized photon is only sensitive to the spin population, and not to the spin coherence.
	Indeed, using our analytical model, one finds that \mbox{$P_{H|M_1}(\tau)=|r_{\downarrow\uparrow}|^2\rho^{\rm (s)}_{|M_1,\downarrow\downarrow}(\tau)$} and $P_{V|M_1}~(\tau)=~|r_{\downarrow\downarrow}|^2\rho^{\rm (s)}_{|M_1,\downarrow\downarrow}(\tau) + |r_{\uparrow\uparrow}|^2\rho^{\rm (s)}_{|M_1,\uparrow\uparrow}(\tau)$\LL{, with no coherence involved}. 
	An $H$-polarized reflected photon can only arise from the Raman spin-flip transition $(H,\omega_2)$, meaning that the spin was in the $\ket{\downarrow}$ state. Conversely, a $V$-polarized reflected photon can be obtained from the resonant $(V,\omega_1)$ transition for spin $\ket\downarrow$, or from the directly reflected laser for spin $\ket\uparrow$. For delays longer than the spin relaxation time, the spin populations relax back to the stationary regime and, as a result, $P_{H|M_1}$ and $P_{V|M_1}$ relax back to the uncorrelated probabilities $P_H$ and $P_V$. The antibunching feature in the $H|M_1$ correlation, i.e. the observation that $P_{H|M_1}(\tau)<~P_H$ at short positive delays, is thus a direct signature that $\rho^{\rm (s)}_{|M_1,\downarrow\downarrow}(\tau)<P_\downarrow$ (and $\rho^{\rm (s)}_{|M_1,\uparrow\uparrow}(\tau)>P_\uparrow$). In turn, this conditional spin population imbalance leads to the observed bunching in the $V|M_1$ correlation, i.e. $P_{V|M_1}(\tau)>P_V$ at short positive delays \cite{SM}.
	
	When $T=D/A, R/L$, the spin coherence induced by the $M_1$-polarized detection (i.e. $\rho_{|M_1,\downarrow\uparrow}^{\rm (s)}(\tau)\neq0$) leads to damped oscillations of the reflected photon polarization, mapping the spin precession. Additional features are also present in all correlations, as the total conditional reflectivity $P_{\mathrm{refl}| M_1}(\tau)$, including both polarizations $T$ and $\bar T$, is also increased at short positive delays, and decreases back to the stationary value as spin relaxation occurs. We discuss this effect in more detail with additional data in the Supplemental Materials \cite{SM}. 
	
	An important interest of the tomography approach is that all the effects are decoupled by focusing on the polarization state of the second photon, regardless of the total reflectivity $P_{\mathrm{refl},|M_1}(\tau)$. This is performed by extracting the conditional Stokes parameters $s_{T\bar{T}|M_1}(\tau)$ as defined by Eq.~(\ref{eq:definition_conditional_stokes_parameters}) and shown in Fig.~2b. In this figure, the parameters $s_{DA|M_1}(\tau)$ and $s_{RL|M_1}(\tau)$ display damped oscillations, following the evolution of the Larmor spin precession around the applied magnetic field. In addition to the spin relaxation, the parameter $s_{HV|M_1}(\tau)$ presents a transient regime for $\tau<1\rm~ns$, leading to an increase. This regime is associated to the time required for the reflection amplitudes to reach their stationary values under a CW drive, which we do not consider in our analytical model. This transient regime also affects the conditional probabilities $P_{T|M_1}(\tau)$, and thus the two-photon correlations in Fig.~2a, for delays shorter than $1~\rm ns$. 
	
	We complement these discussions with a quantitative analysis through the physical parameters extracted from a fully numerical model of the QD-cavity system \cite{Mehdi2024,Gundin2025, SM}. This model allows reproducing all the experimental data discussed in this paper, including additional measurements displayed in Supplemental Materials \cite{SM}. It includes the light-matter coupling strength $g = 2\pi\times(3.1 \pm 0.2)~\rm GHz$ and the spontaneous emission rate outside the cavity mode $\gamma_{\mathrm{sp}}=2\pi\times(157\pm13)~\rm MHz$, together with an homogeneous pure dephasing of the radiative transitions, $\gamma^*=2\pi\times(24\pm 12)~\rm MHz$. Together with the cavity parameters, the QD parameters $g$, $\gamma_{\mathrm{sp}}$ and $\gamma^*$ determine the device optical response in polarisation, and thus the amount of measurement back-action induced by photon detection. In addition, the parameters $g$ and $\gamma_{\mathrm{sp}}$ determine the radiative lifetime $T_{\rm rad}= 470\pm 30\rm~ps$~\cite{SM}, which in turn governs the transient regime visible at short delays. 
	
	Our numerical simulations also include charge dynamics, as described by the charge escape time $\tau_{\rm esc}=4.1\pm0.2~\rm ns$ and the charge occupation probability $P_{\rm c} = 0.96 \pm 0.02$ \cite{SM}. Indeed the device operates in a co-tunneling regime between the QD and the nearby \mbox{n-doped} region \cite{Mehdi2024}, leading to frequent spin resets which determine the spin relaxation time, $T_{1,\rm spin}=\tau_{\rm esc}$. 
	
	Finally, our simulations include the electron transverse Landé factor $g_{e,\perp}=0.48\pm0.01$ which governs the Larmor precession with period \mbox{$T_L=\frac{h}{g_{e,\perp}\mu_BB}=745\pm 15\rm~ps$}. We also take into accout the interaction between the electron and the Overhauser field \cite{Urbaszek2013} with coupling strength \mbox{$\gamma_e=2\pi\times(120\pm12)~\rm MHz$} \cite{SM}, which governs the inhomogeneous spin coherence time $T_{2,\rm spin}^*=\sqrt{2}
	/\gamma_e=1.9\pm0.2\rm~ ns$ \cite{SM,Coste2023}. \\
	\begin{figure}[t!]
		\includegraphics[scale=1]{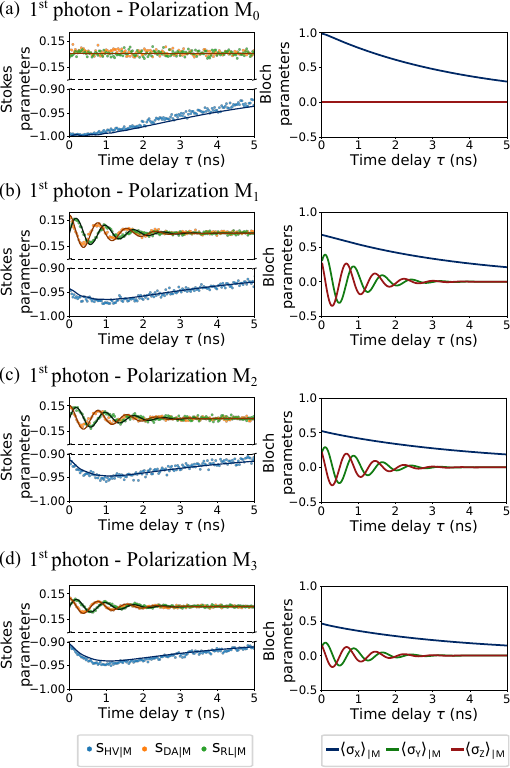}
		\caption{Time-dependence of the measured Stokes parameters (Left) and the simulated spin state (Right) as a function of the polarization state used to measure the first photon. (a) Polarization $M_0$, (b) Polarization $M_1$, (c) Polarization $M_2$, (d) Polarization 
			$M_3$. \LL{In the left panels the scatter plot display the experimental data and the black solid lines correspond to numerical simulations.
		}} 
		\centering
	\end{figure}
	
	\subsection{Measurement-induced quantum back-action}
	Our numerical model, associated with the full tomography of the second photon's polarization, enables us to study the effect that the first photon detection has on the spin.
	Fig.~3 displays both the measured reflected photon state (Stokes parameters $s_{T\bar{T}|M}$, left panels) and the simulated conditional spin state (Bloch parameters $\left<\hat{\sigma}_i \right>_{|M}$ with $\hat{\sigma}_i$ the spin Pauli matrix in the  $i=x,y,z$ axis, right panels),
	for the polarization states (a) $M_0$, (b) $M_1$, (c) $M_2$ and (d) $M_3$. We note that the data in the left panel of Fig.~3b is the same as in Fig.~2b. When the first photon is measured on the $M_0=H$ polarization, the spin is projected into the $\ket{\uparrow}$ eigenstate, \LL{as shown in the right panel of Fig. 3a}. In this situation, the back-action induced by the photon detection is classical, modifying the spin state populations with no induced coherence. 
	\LL{As expected from our analytical model, the use of of a different measurement polarization angle, $\phi\neq 0$, allows projecting} the spin in a partially coherent superposition of its energy eigenstates. The spin then precesses around the magnetic field\LL{, as shown  in the right panels of Fig. 3b} for $M_1$, $M_2$ and $M_3$.
	The choice of polarization determines the amplitude of the oscillations of $\left<\hat{\sigma}_y \right>_{|M}$ and $\left<\hat{\sigma}_z \right>_{|M}$, associated to the real and imaginary parts of $\rho^{\mathrm{(s)}}_{|M, \downarrow\uparrow}$, as expected from Eq.~(\ref{eq:state_after_measurement}). This contrast variation is then transferred onto the second photon polarization (see Eq.~(\ref{eq:conditional_stokes_parameters})).
	
	Finally, we benchmark our analytical model against our experimental data and numerical simulations. First, we infer the occupation probabilities in the stationary regime from the numerical model: $P_\uparrow=0.51$ and \mbox{$P_\downarrow=0.49$}, where the difference from 0.5 arises from the residual optical spin pumping induced by the excitation of the $(V,\omega_1)$ transition. Then, we extract the unconditional reflection probabilities in polarizations $H$ and $V$, under $V$ polarized excitation, which we denote as $P^{\mathrm{(avg)}}_{V\rightarrow V}$ and $P^{\mathrm{(avg)}}_{V\rightarrow H}$ \cite{Mehdi2024}. The results are displayed in Fig.~4a, together with their measured values, as a function of the detuning between the laser and the QD frequency at 0T. We infer the reflectivity coefficients $|r_{\downarrow\downarrow}|^2=0.37$ and $|r_{\downarrow\uparrow}|^2=0.06$ from the simulated values of the reflection probabilities at $\omega_{\rm laser}=\omega_1$ and of the populations $P_\uparrow$ and $P_\downarrow$ \cite{SM}.
	In addition, $|r_{\uparrow\uparrow}|^2$ is directly taken as $|r_{\uparrow\uparrow}|^2=P^{\mathrm{(cav)}}_{V\rightarrow V}=0.69$, the empty cavity reflectivity response in polarization $V$. These reflection probabilities are also displayed in Fig.~4a for comparison with the average reflectivities.
	
	We assess the amount of coherence induced onto the spin by the photon detection using the simulated values of $P_\uparrow$, $P_\downarrow$, $|r_{\uparrow\uparrow}|$, $|r_{\uparrow\downarrow}|$ and $|r_{\downarrow\downarrow}|$ in Eqs. (5) to (9). We define the Bloch (Stokes) coherence $C_B$ ($C_S$) as the amplitude of oscillations of $\left<\sigma_y\right>_{|M}$ and $\left<\sigma_z\right>_{|M}$ ($s_{DA|M}$ and $s_{RL|M}$) at $\tau=0$ :
	\begin{align}
		C_B(\phi) &= \sqrt{ \left\langle \sigma_y \right\rangle_{|M}^2(0) + \left\langle \sigma_z \right\rangle_{|M}^2(0)} \\
		C_S(\phi) &= \sqrt{s^2_{DA|M}(0) + s^2_{RL|M}(0)}
	\end{align}
	\noindent as a function of the angle of the detected photon's polarization state. Since the Bloch coherence can also be rewritten as $C_B = 2|\rho^{\mathrm{(s)}}_{|M, \downarrow\uparrow}|$, it provides a measurement of the off-diagonal elements of the spin density matrix after detection, and  therefore quantifies the amount of spin coherence. The Stokes coherence is also proportional to the Bloch coherence as $C_S=2|\rho^{\mathrm{(s)}}_{|M, \downarrow\uparrow}r_{\uparrow\uparrow}^*r_{\downarrow\uparrow}|/P_{\mathrm{refl}|M}$.
	Fig.~4b shows the experimentally measured $C_S$ (red data points) as well as numerical simulations for both $C_S$ and $C_B$ (solid lines). As expected from our model, the coherences reach zero at $\phi=0\degree$ and $\phi=180^\circ$. In addition, based on the reflection amplitudes which govern $\rho_{|M,\downarrow\uparrow}^{\rm (s)}$, an optimal angle is obtained when $\phi$ is close to $30^\circ$, i.e. close to polarization $M_1$. Fig.~4b also displays the  calculated coherences (dashed lines), using our analytical model, which qualitatively reproduces all these features. The discrepancy between the numerical and the analytical model comes from the latter not considering environmental noise sources (pure dephasing, hyperfine coupling, electron cotunneling) nor the transient regime associated to the finite  radiative lifetime $T_\mathrm{rad}$.\\

	\begin{figure}
		\includegraphics[scale=1.0]{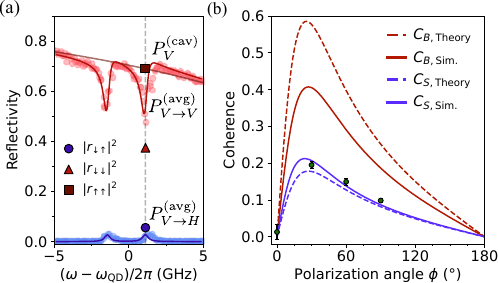}
		\caption{(a) Average polarization-dependent intensities for a QD in the thermal state as a function of the laser-QD detuning under $V$-polarized excitation for a charged (red, blue) and uncharged QD (brown). The $|r_{\downarrow\downarrow}|^2$ and $|r_{\downarrow\uparrow}|^2$ spin-flip coefficients of (\ref{interaction_description}) are extracted from these data (see main text). (b) Experimentally measured Stokes coherence (red dots) as a function of the measurement angle $\phi$ compared with numerical simulations (solid line) and the analytical description (dashed line).
		}
		\centering
	\end{figure}
	
	\section{CONCLUSION}
	In this work, we have introduced a time-resolved tomography approach to measure the dynamics of a single spin state, mapped to the polarization state of a single reflected photon. Our results allow evidencing, quantifying and controlling the quantum back-action induced on the spin system through the detection of a single reflected photon. We introduced an analytical model qualitatively explaining the experimental data, allowing us to predict the conditional spin state after photon detection, and the conditional polarization state for a second reflected photon. We find that the spin dynamics, fully described by the conditional Bloch components $\left<\hat{\sigma}_x \right>_{|M}$, $\left<\hat{\sigma}_y \right>_{|M}$ and $\left<\hat{\sigma}_z \right>_{|M}$, is in correspondence with the polarization dynamics of the second reflected photon, as described by the conditional Stokes coefficients $s_{HV|M}(\tau)$, $s_{DA|M}(\tau)$ and $s_{RL|M}(\tau)$. Therefore, with a single \LL{time-resolved, conditional} tomography measurement, one can extract all the timescales of the spin dynamics and quantify the back-action induced by the detection of a single reflected photon.
	The observed results are in complete agreement with numerical simulations  of the QD-cavity system \cite{Mehdi2024,Gundin2025} which, contrary to the analytical model, take into account the environmental noise sources and the finite trion lifetime. 
	
	A first perspective for this work lies in the possibility to induce a maximally-coherent spin precession ($C_B=1$, instead of $C_B \approx 0.4$ here), ensuring that the conditional spin state is a pure state oscillating as $\cos\left(\pi\frac{t}{T_\mathrm{L}}\right) \ket{\uparrow}_z+i \sin\left(\pi\frac{t}{T_\mathrm{L}}\right)\ket{\downarrow}_z$. This is a prerequisite to ensure that a second photon, incoming at a specific time such as $t=\frac{T_\mathrm{L}}{4}$, becomes maximally entangled with the precessing spin \cite{Hu2008,Hu2008_Entangler}. This will require operating in the pulsed regime, using devices with optimized Purcell enhancements (reducing the QD-cavity detuning) and increased spin coherence \cite{Coste2023,Hogg2025}. This could allow producing one-dimensional photonic cluster states through the Lindner-Rudolf protocol \cite{Lindner2009}, in a potentially deterministic way, using incoming/reflected photons instead of emitted ones. The final goal will be the use of spin-photon interfaces in quantum communication and quantum computing schemes exploiting a few nodes \cite{Borregaard2020}, and potentially a single node used both as emitter and receiver \cite{Pichler2017,Zhan2020,Ferreira2024}, to produce multidimensional photonic cluster states.\\
	
	\section{Appendix}
	\subsection{Device fabrication}
	The QD-cavity device was grown by molecular beam epitaxy and consists of a $\lambda$-GaAs cavity, formed by two distributed Bragg reflectors, embedding an annealed InGaAs QD. The Bragg mirrors are made by alternating layers of GaAs and Al$_{0.9}$Ga$_{0.1}$As, with 20 (30) pairs for the top (bottom) mirror. To electrically contact the structure, the bottom mirror (Si-doped) presents a gradual doping from $2\times10^{18}$~cm$^{-3}$ down to $1\times10^{18}$~cm$^{-3}$. This level of doping is maintained in the first half of the cavity region and is stopped only 25~nm before the QD layer, which creates a tunnel barrier between the quantum dot and the Fermi sea. The top mirror is C-doped with increasing doping level, from zero to $2\times10^{19}$~cm$^{-3}$ at the surface. The micropillar is connected through four ridges to a large circular frame, attached to a gold-plated mesa enabling the electrical control. For full details on the device fabrication, see \cite{Somaschi2016,DeSantis2017}.\\

	\textbf{ACKNOWLEDGEMENTS} \\
	This work was partially supported by the Paris Ile-de-France Région in the framework of DIM SIRTEQ, the Research and Innovation Programme QUDOT-TECH under the Marie Sklodowska-Curie grant agreement 861097, Horizon CL4 program under the
	grant agreement 101135288 for EPIQUE project, the European Union’s Horizon 2020 FET OPEN project QLUSTER (Grant ID 862035), the French National Research Agency (ANR) through project ANR-22-PETQ0013, a public grant as part of the ”Investissements d’Avenir” programme (Labex NanoSaclay, reference: ANR10LABX0035) and a government grant as part of France 2030 (QuanTEdu-France), under reference ANR-22-CMAS-0001. This work was done within the C2N micro nanotechnologies platforms and partly supported by the RENATECH network and the General Council of Essonne.\\ 
	

	\textbf{COMPETING INTERESTS} \\
	N.S. is a co-founder and P.S. is a scientific advisor and co-founder of the company Quandela. The other authors declare no competing interests.

	\bibliographystyle{naturemag_noURL}


\end{document}